\documentclass[prd,aps,showpacs,,nofootinbib,superscriptaddress,preprintnumbers,epsf,psf]{revtex4}

\usepackage[english]{babel}
\usepackage{pstricks}
\usepackage[dvips]{graphicx}
\usepackage{epsf}
\usepackage{amsmath}
\usepackage{amsfonts}
\usepackage{amssymb}

\begin{document}

\title{Gauge invariance, causality and gluonic poles}

\author{I.~V.~Anikin}
\email{anikin@theor.jinr.ru}
\affiliation{Bogoliubov Laboratory of Theoretical Physics, JINR,
             141980 Dubna, Russia}
\author{O.~V.~Teryaev}
\email{teryaev@theor.jinr.ru}
\affiliation{Bogoliubov Laboratory of Theoretical Physics, JINR,
            141980 Dubna, Russia}

\begin{abstract}
We explore the electromagnetic gauge invariance of the hadron tensor of
the Drell-Yan process with one transversely polarized hadron.
The special role is played by the contour gauge for gluon fields.
The prescription for the gluonic pole in the twist $3$ correlator is related to
causality property and compared with the prescriptions for exclusive hard processes.
As a result we get the extra contributions, which naively
do not have an imaginary phase. The single spin asymmetry for
the Drell-Yan process is accordingly enhanced by the factor of two.
\end{abstract}
\date{\today}
\maketitle

\section{Introduction}
\label{Introduction}

Recently, the problem of the electromagnetic gauge invariance in the
deeply virtual Compton scattering (DVCS) and similar exclusive processes
has intensively been discussed in the literature, see for example
\cite{Gui98, Pire, APT-GI, Bel-Mul, PPSS}.
This development explored the similarity with the earlier studied inclusive spin-dependent
processes \cite{Efr}, and the transverse component of momentum transfer in DVCS
corresponds to the transverse spin in DIS.

The gauge invariance of relevant amplitudes is
ensured by means of twist three contributions and the use of the
equations of motion providing a possibility to
exclude the three-particle (quark-gluon) correlators from the amplitude.
After combining with the two-particle
correlator contributions, one get the
gauge invariant expression for the physical amplitude or, in the case of lepton-hadron processes,
 for the corresponding hadron
tensor \cite{Efr}.

This method was originally developed in the case of the particular inclusive processes
with transverse polarized hadrons, like structure function $g_2$ in 
DIS \cite{Efr} and Single Spin Asymmetry (SSA)
 \cite{EKT} due to soft quark
(fermionic poles \cite{Efremov:1984ip}).
At the same time, the colour gauge invariance of the so-called gluonic poles contributions \cite{b}
was previously explored \cite{Boer:2003cm} by other methods  relying on the Wilson exponentials
\cite{Collins:2002kn,Belitsky:2002sm,Collins:2004nx,Cherednikov:2008ua}.

Here we combine the approaches described above and apply them in the
relevant case of the Drell-Yan (DY) process where one of hadrons is the transversally 
polarized nucleon.

The SSA in the DY process was first considered in QCD in the case \cite{P-R,Carlitz:1992fv} of 
the longitudinally polarized hadron. This observable is especially 
interesting if the second hadron is a pion, because of the sensitivity  
\cite{Brandenburg:1995pk, Bakulev:2007ej}
to the shape of pion distribution amplitude, being currently the object of major interest 
\cite{Radyushkin:2009zg,Polyakov:2009je}
(see also \cite{Mikhailov:2009sa} and Ref. therein).     

The imaginary phases in the SSA with longitudinally polarized nucleon 
are due to the hard perturbative gluon loops \cite{P-R,Carlitz:1992fv} 
or twist $4$ contribution of the  pion distribution amplitude 
\cite{Brandenburg:1994wf,Brandenburg:1995pk,Bakulev:2007ej}. At the same time,  
the source of the imaginary part, when one
calculates the single spin asymmetry associated with
$P+P^{\uparrow\downarrow}\to\ell \bar \ell + X$ process, is the quark propagator in
the diagrams with quark-gluon (twist three) correlators. This leads \cite{Teryaev} to
the gluonic pole contribution to SSA.
It has been reproduced (up to the derivative term, corresponding to the case of single 
inclusive Drell-Yan process, 
when only one of the leptons is observed) in the case of the non-zero boundary
condition imposed on gluon fields, and the asymmetric boundary conditions
have been considered as a privileged ones \cite{Boer}.
The reason is that these boundary conditions provide the purely real
quark-gluon function $B^V(x_1,x_2)$ which parameterizes
$\langle\bar\psi\gamma^+A_\alpha^T\psi\rangle$ matrix element.
By this fact the diagrams with two-particle correlators do not contribute to the
imaginary part of the hadron tensor related to the SSA.
This property seems quite natural, as the corresponding diagram does not have
a cut capable of producing the imaginary phase \cite{Ter00}.

In our paper, we perform a thorough analysis of the transverse polarized DY hadron tensor
in the light of the QED gauge invariance, the causality and gluonic pole contributions.

We show that to restore the electromagnetic gauge invariance of the
transverse polarized DY hadron tensor, it is mandatory to add the extra diagram contribution 
(cf. \cite{BQ}), also at the twist three level.
In contrast to the naive assumption,
we demonstrate that this new additional contribution is directly related to
the certain complex prescription in the gluonic pole $1/(x_1-x_2)$ of
the quark-gluon function $B^V(x_1,x_2)$. 
It is essential that this
prescription is process-dependent, supporting the idea of effective process-dependent
Sivers function (see e.g.\cite{c} and Refs. therein) related to this correlator.

In more detail, we show that the causal pole prescription in the quark propagator, 
involved in the hard part of
the standard diagram, supports  the choice of a contour gauge and, in turn,
the representation of the quark-gluon function $B^V(x_1,x_2)$ in
the form of the gluonic pole with the mentioned complex prescription.
This representation
must be extended on the diagram, which naively does not contribute to the imaginary part.
They ensure a new contribution to the imaginary part which is necessary 
to maintain the electromagnetic gauge invariance.
Finally,
the account for this new contributions corrects  the SSA formula for
the transverse polarized Drell-Yan process by the factor of $2$.

\section{Causality and contour gauge for the gluonic pole}
\label{Causality}

We study the contribution to the hadron tensor which is related to the single spin
(left-right) asymmetry
measured in the Drell-Yan process with the transversely polarized nucleon:
$N^{(\uparrow\downarrow)}(p_1) + N(p_2) \to \gamma^*(q) + X(P_X)\to
\ell(l_1) + \bar\ell(l_2) + X(P_X)$,
where the virtual photon producing the lepton pair ($l_1+l_2=q$) has a large mass squared 
($q^2=Q^2$)
while the transverse momenta are small and integrated out.
The left-right asymmetry means that the transverse momenta
of the leptons are correlated with the direction 
$\textbf{S}\times \textbf{e}_z$ where $S_\mu$ implies the
transverse polarization vector of the nucleon while $\textbf{e}_z$ is a beam direction \cite{Barone}.

The DY process with the transversely polarized target
manifests \cite{Teryaev} the gluonic pole contributions.
Since we perform our calculations within a {\it collinear} factorization,
it is convenient (see,e.g., \cite{An})
to fix  the dominant light-cone directions for the DY process
shown at Fig. \ref{Fig-DY}
\begin{eqnarray}
\label{kin-DY}
p_1\approx \frac{Q}{x_B \sqrt{2}}\, n^*\, , \quad p_2\approx \frac{Q}{y_B \sqrt{2}}\, n\,\,\,\,
\text{with} \,\,\,\,
n^*_{\mu}=(1/\sqrt{2},\,{\bf 0}_T,\,1/\sqrt{2}), \quad n_{\mu}=(1/\sqrt{2},\,{\bf 0}_T,\,-1/\sqrt{2})\, ,
\end{eqnarray}
so that the hadron momenta $p_1$ and $p_2$ have the plus and minus dominant light-cone
components, respectively. Accordingly, the quark and gluon momenta $k_1$ and $\ell$ lie
along the plus direction while the antiquark momentum $k_2$ -- along the minus direction.

Focusing on the Dirac vector projection, containing the gluonic pole,
let us start with the standard hadron tensor
generated by the diagram depicted on Fig. \ref{Fig-DY}(a):
\begin{eqnarray}
\label{HadTen1-2}
{\cal W}^{(1)}_{\mu\nu}&=& \int d^4 k_1\, d^4 k_2 \, \delta^{(4)}(k_1+k_2-q)
\int d^4 \ell \,
\Phi^{(A)\,[\gamma^+]}_\alpha (k_1,\ell) \, \bar\Phi^{[\gamma^-]} (k_2)\times
\nonumber\\
&&\text{tr}\biggl[
\gamma_\mu  \gamma^- \gamma_\nu \gamma^+ \gamma_\alpha
\frac{\ell^+\gamma^- - k_2^-\gamma^+}
{-2\ell^+ k_2^- + i\epsilon}
\biggr] \, ,
\end{eqnarray}
where
\begin{eqnarray}
\label{PhiF}
\Phi^{(A)\,[\gamma^+]}_\alpha (k_1,\ell)
\stackrel{{\cal F}_2}{=}
\langle p_1, S^T | \bar\psi(\eta_1)\gamma^+  gA_{\alpha}(z)  \psi(0) | S^T, p_1\rangle ,
\quad
\bar\Phi^{[\gamma^-]}(k_2)\stackrel{{\cal F}_1}{=}
\langle p_2 | \bar\psi(\eta_2)\gamma^- \psi(0)| p_2\rangle .
\end{eqnarray}

Throughout this paper, ${\cal F}_1$ and  ${\cal F}_2$
denote the Fourier transformation with the measures
\begin{eqnarray}
d^4\eta_2\, e^{ik_2\cdot\eta_2}\,\,\, \text{and} \,\,\,
d^4\eta_1\, d^4 z\, e^{-ik_1\cdot\eta_1-i\ell\cdot z} ,
\end{eqnarray}
respectively, while ${\cal F}_1^{-1}$ and ${\cal F}_2^{-1}$ mark the inverse
Fourier transformation with the measures
\begin{eqnarray}
dy \, e^{i y\lambda}\,\,\, \text{and} \,\,\,
dx_1 dx_2 \, e^{i x_1\lambda_1+ i(x_2 - x_1)\lambda_2}.
\end{eqnarray}
Analyzing the $\gamma$-structure of (\ref{HadTen1-2}), we may conclude that
the first term in the quark propagator singles out the combination:
$\gamma^+ \gamma_\alpha \gamma^-$ with $\alpha=T$ which
will lead to the matrix element of the twist three operator,
$\langle \bar\psi\, \gamma^+ A^T_\alpha \psi\rangle$ with
the transverse gluon field. After factorization, this matrix element
will be parametrized via the function $B^V(x_1,x_2)$.
The second term in the numerator of the quark propagator separates out
the combination  $\gamma^+ \gamma_\alpha \gamma^+$ with $\alpha=-$.
Therefore, this term will give $\langle \bar\psi\, \gamma^+\,  A^+\,\psi\rangle$
which, as we will see now, will be exponentiated in the Wilson line
$[-\infty^-,\, 0^-]$.
Indeed, this part of the standard hadron tensor is given by
\begin{eqnarray}
\label{HadTen1-3}
{\cal W}^{(1)\, [k_2^- -\text{term}]}_{\mu\nu}&=& \int d\mu (k_i;x_1,y) \,
\text{tr}\biggl[
\gamma_\mu  \gamma^- \gamma_\nu \gamma^+ \gamma^- \gamma^+
\biggr] \, \bar\Phi^{[\gamma^-]} (k_2) \times
\\
&&\frac{1}{2} \int dz^- \int d\ell^+\, \frac{e^{ - i\ell^+ z^-}}{\ell^+ - i\epsilon}
\int d^4\eta_1  \, e^{-ik_1\cdot\eta_1}
\langle p_1, S^T | \bar\psi(\eta_1)\, \gamma^+ \, gA^+(0,z^-,\vec{{\bf 0}}_T) \, \psi(0)
|S^T, p_1\rangle \, ,
\nonumber
\end{eqnarray}
where 
\begin{eqnarray}
d\mu (k_i;x_1,y)= dx_1 d^4k_1 \delta(x_1-k_1^+/p_1^+)\,dy d^4k_2 \delta(y-k_2^-/p_2^-)\,
\biggl[ \delta^{(4)}(x_1p_1+yp_2 - q) \biggr] \, .
\end{eqnarray}
Note that the prescription $-i\epsilon$ in the denominator of this expression
directly follows from the standard (see, e.g. \cite{BogoShir}) causal 
prescription for the massless quark propagator in (\ref{HadTen1-2}).

Integrating over $\ell^+$, one can immediately obtain
the corresponding $\theta$-function in (\ref{HadTen1-3}):
\begin{eqnarray}
\label{HadTen1-4}
{\cal W}^{(1)\,  [k_2^- -\text{term}]}_{\mu\nu}&=& \int d\mu (k_i;x_1,y) \,
\text{tr}\biggl[
\gamma_\mu  \gamma^- \gamma_\nu \gamma^+
\biggr] \, \bar\Phi^{[\gamma^-]} (k_2) \times
\nonumber\\
&&
\int d^4\eta_1 \, e^{-ik_1\cdot\eta_1}
\langle p_1, S^T | \bar\psi(\eta_1)\, \gamma^+ \,
ig \int\limits_{-\infty}^{+\infty} dz^- \, \theta(-z^-) A^+(0,z^-,\vec{{\bf 0}}_T)
\, \psi(0) |S^T, p_1\rangle \, .
\end{eqnarray}
Including all gluon emissions from the antiquark going from the upper blob
on Fig. \ref{Fig-DY}(a) (the so-called initial state interactions),
we get the corresponding $P$-exponential in
$\Phi^{(A)\,[\gamma^+]}_\alpha (k_1,\ell)$. The latter is now represented by the
following matrix element:
\begin{eqnarray}
\label{me-Pexp}
\int d^4\eta_1 \, e^{-ik_1\cdot\eta_1}
\langle p_1, S^T | \bar\psi(\eta_1)\, \gamma^+ \, [-\infty^-,\, 0^-]
\, \psi(0) |S^T, p_1\rangle \, ,
\end{eqnarray}
where
\begin{eqnarray}
\label{Pexp-1}
[-\infty^-,\, 0^-] = Pexp\biggl\{ - i g \int\limits_{-\infty}^{0}
 dz^- \, A^+(0,z^-,\vec{{\bf 0}}_T) \biggr\}\, .
\end{eqnarray}
If we include in the consideration the gluon emission from
the incoming antiquark (the mirror contributions), we will obtain the Wilson
line $[\eta_1^-,-\infty^-]$ which
will ultimately give us, together with (\ref{Pexp-1}), the Wilson line connecting
the points $0$ and $\eta_1$ in (\ref{me-Pexp}).
This is exactly what happens, say, in the spin-averaged DY process \cite{Efremov:1978xm}.
However, for the SSA, these
two diagrams should be considered individually.
Indeed, their contributions to SSAs, contrary to spin-averaged case,
differ in sign and the dependence on the boundary point at $-\infty^-$
does not cancel.

To eliminate the unphysical gluons from our consideration and use the 
factorization scheme \cite{Efr},
we may choose a {\it contour} gauge \cite{ContourG}
\begin{eqnarray}
\label{cg2}
[-\infty^-,\, 0^-]=1 \,
\end{eqnarray}
which actually implies also the axial gauge $A^+=0$ used in \cite{Efr}.

Let us discuss the problem of gauge choice in more detail.
In (\ref{cg2}), the so-called starting point $x_0$ (see, \cite{ContourG}) is fixed to be 
at $-\infty^-$ owing
to the certain complex prescription $+i\epsilon$ in the quark propagator in (\ref{HadTen1-2}).
If we would change the starting point $x_0$ on $+\infty^-$, this would correspond to the choice
of the ''anticausal" complex prescription $-i\epsilon$.
On the other hand, the axial gauge $A^+=0$ is independent on the choice of $x_0$ and 
we are able to eliminate
the Wilson line by choosing simply $A^+=0$ without referring to the starting point $x_0$.
Nevertheless, since our prescription $+i\epsilon$ in the quark propagator uniquely fixes 
the starting point $x_0$ at $-\infty$, the expression for the Wilson line 
(\ref{Pexp-1}) hints the choice of gauge (\ref{cg2}).

Imposing this gauge one arrives \cite{ContourG} at the following representation 
of the gluon field in terms of the strength tensor:
\begin{eqnarray}
\label{Ag}
A^\mu(z)=
\int\limits_{-\infty}^{\infty} d\omega^- \theta(z^- - \omega^-) G^{+\mu} (\omega^-)
+ A^\mu(-\infty) \, .
\end{eqnarray}
Moreover, as we will demonstrate below, if we choose instead an
alternative representation for the gluon in the form:
\begin{eqnarray}
\label{AgAlter}
A^\mu(z)=
- \int\limits_{-\infty}^{\infty} d\omega^- \theta( \omega^- - z^- ) G^{+\mu} (\omega^-)
+ A^\mu(\infty) \,
\end{eqnarray}
(which corresponds to the gauge condition $[+\infty^-,\, 0^-]=1$ and also results in $A^+=0$)
keeping the causal prescription
$+i\epsilon$ in (\ref{HadTen1-2}), the cost of this will be the breaking of the electromagnetic gauge
invariance for the DY tensor.

We are now ready to pass to the term with $\ell^+\gamma^-$ in (\ref{HadTen1-2})
which gives us finally the matrix element of the twist
three quark-gluon operator with the transverse gluon field.
Let us stop, in more detail, on the parametrization of the relevant matrix elements:
\begin{eqnarray}
\label{parVecDY}
&&\langle p_1, S^T | \bar\psi(\lambda_1 \tilde n)\, \gamma_{\beta} \,
g A_{\alpha}^T(\lambda_2\tilde n) \,\psi(0)
|S^T, p_1 \rangle
\stackrel{{\cal F}_2^{-1}}{=}
i\varepsilon_{\beta\alpha S^T p_1} \, B^V(x_1,x_2)\, .
\end{eqnarray}
Using the representation (\ref{Ag}), this function can be expressed as
\begin{eqnarray}
\label{Sol-way-1}
B^V(x_1,x_2)= \frac{T(x_1,x_2)}{x_1-x_2+i\epsilon} + \delta(x_1-x_2) B^V_{A(-\infty)}(x_1)\, ,
\end{eqnarray}
where the real regular function $T(x_1,x_2)$ ( $T(x,x) \neq 0$)
parametrizes
the vector matrix element of the operator involving the tensor $G_{\mu\nu}$ (cf. \cite{An-ImF}):
\begin{eqnarray}
\label{parT}
\langle p_1, S^T | \bar\psi(\lambda_1 \tilde n)\, \gamma_{\beta} \,
\tilde n_\nu G_{\nu\alpha}(\lambda_2\tilde n) \,\psi(0)
|S^T, p_1 \rangle\stackrel{{\cal F}^{-1}_2}{=}
\varepsilon_{\beta\alpha S^T p_1}\,
T(x_1,x_2)\, .
\end{eqnarray}
Owing to the time-reversal invariance, the function $B^V_{A(-\infty)}(x_1)$,
\begin{eqnarray}
\label{BatInfty}
i\varepsilon_{\beta\alpha S^T p_1} \,\delta(x_1-x_2) B^V_{A(\pm\infty)}(x_1) \stackrel{{\cal F}}{=}
\langle p_1, S^T | \bar\psi(\lambda_1 \tilde n)\, \gamma_\beta \,
g A_{\alpha}^T(\pm\infty) \,\psi(0) | S^T, p_1 \rangle\, ,
\end{eqnarray}
can be chosen as
\begin{eqnarray}
\label{BC-2}
B^V_{A(-\infty)}(x)= 0\, .
\end{eqnarray}
Indeed, the function $B^V(x_1,x_2)$ is an antisymmetric function of its arguments \cite{Efr},
while the anti-symmetrization of the additional term with $B^V_{A(-\infty)}(x_1)$ gives zero.

There is no doubt that
the only source of the imaginary part of the hadron tensor
is the quark propagator.  One may try to realize this property by assumption that matrix elements
are purely real,
\begin{eqnarray}
\label{g-pole-B}
B^V(x_1,x_2)= \frac{{\cal P}}{x_1-x_2} T(x_1,x_2)\, ,
\end{eqnarray}
corresponding to
asymmetric boundary condition for gluons \cite{Boer}:
\begin{eqnarray}
\label{BC-1}
&&B^V_{A(\infty)}(x) = - B^V_{A(-\infty)}(x)
\end{eqnarray}

Here we suggest another way of reasoning. The causal prescription for the quark propagator,
generating its imaginary part, simultaneously leads to the imaginary part of the gluonic pole.
Let us emphasize that this does not mean the appearance of imaginary part of matrix element
(which by itself does not have
an explicit physical meaning) but rather the prescription of its convolution with hard part.
This procedure is in agreement with the prescriptions which were appeared in the
exclusive case in the parametrization of the generalized gluon distributions \cite{Rad, Braun}.

Note that the fixed complex prescription $+i\epsilon$ in the
gluonic pole of $B^V(x_1,x_2)$ (see, (\ref{Sol-way-1})) is one of our main results and
is very crucial for a new contribution to hadron tensor we are now ready to explore.
Indeed, the gauge condition must be the same for all the diagrams, and
it leads to the appearance of imaginary phase of the diagram (see, Fig. \ref{Fig-DY}(b))
which naively does not have it. Let us confirm this by explicit calculation.

\section{Hadron tensor and gauge invariance}
\label{New-tensor}

\begin{figure}[t]
\centerline{\includegraphics[width=0.3\textwidth]{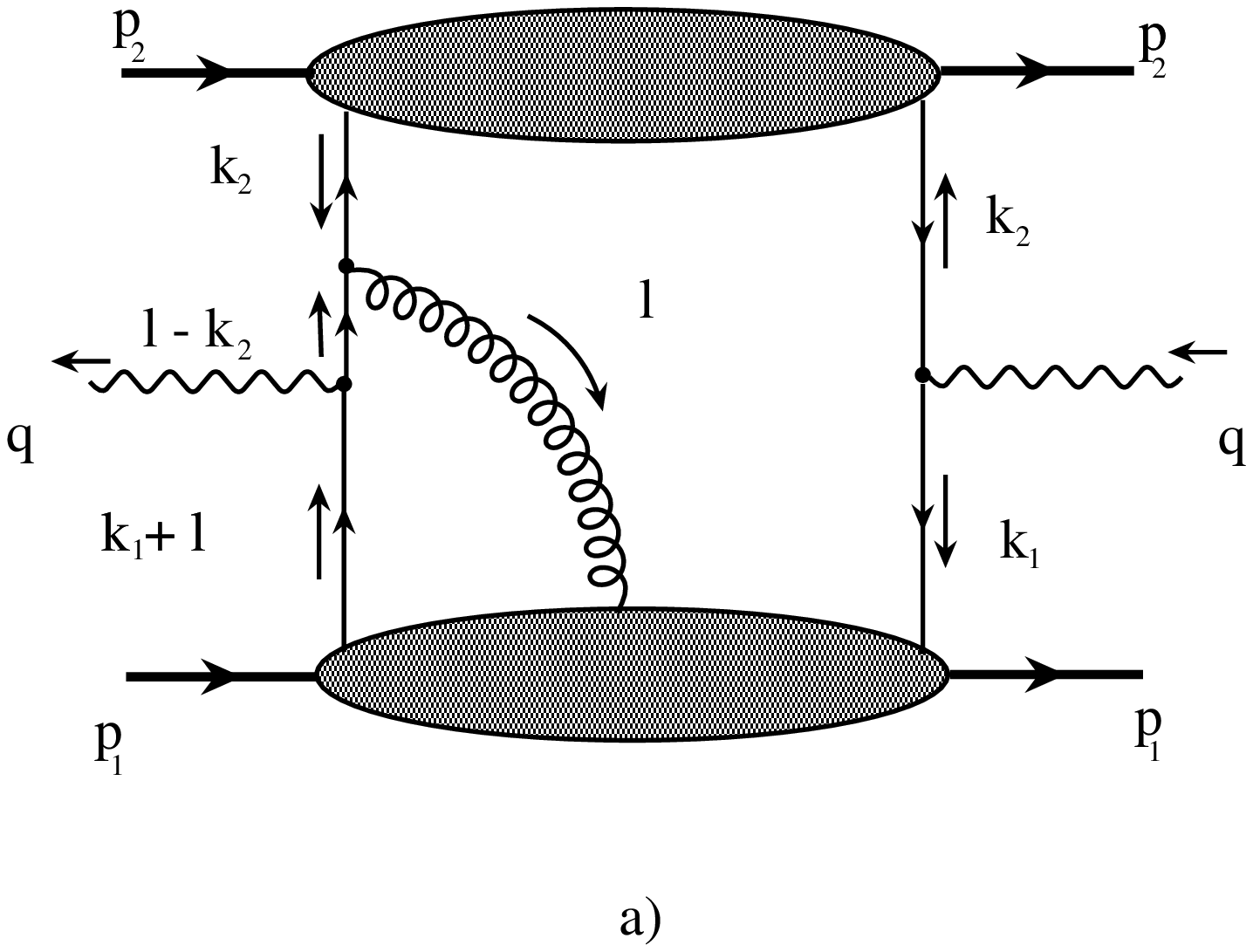}
\hspace{1.cm}\includegraphics[width=0.3\textwidth]{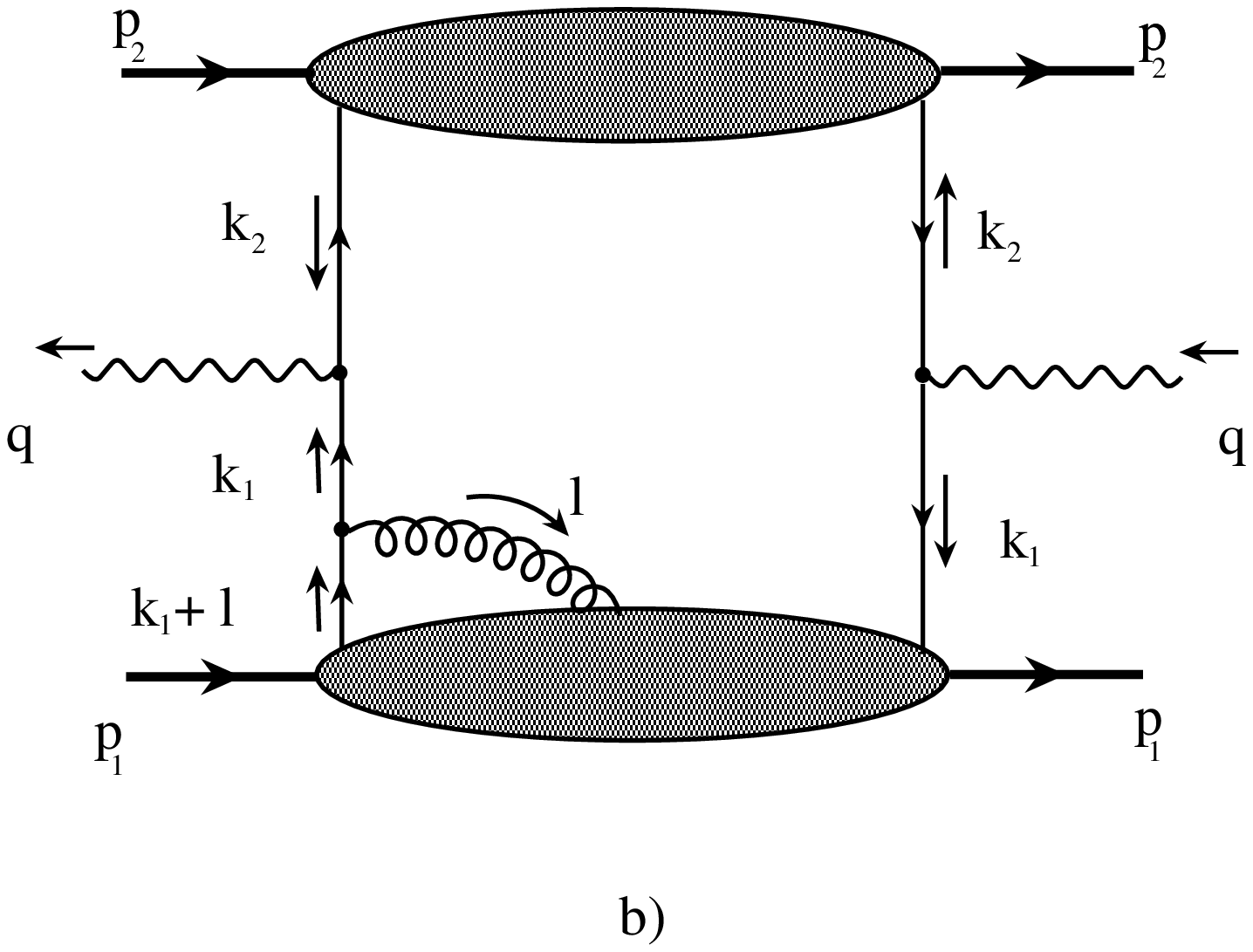}}
  \caption{The Feynman diagrams which contribute to the polarized Drell-Yan hadron tensor.}
\label{Fig-DY}
\end{figure}

We now return to the hadron tensor and calculate the part involving $\ell^+\gamma^-$,
obtaining the following expression for
the standard hadron tensor (see, the diagram on Fig. \ref{Fig-DY}(a)):
\begin{eqnarray}
\label{FacHadTen1}
&&\overline{\cal W}^{(1)\,[\ell^+-\text{term}]}_{\mu\nu}=
\int d^2 \vec{\textbf{q}}_T \, {\cal W}^{(1)}_{\mu\nu}=
- \int dx_1 \, dy \,
\biggl[\delta(x_1-x_B) \delta(y-y_B)\biggr] \, \bar q(y) \times
\\
&& \Im\text{m}\, \int dx_2 \, \text{tr}\biggl[
\gamma_\mu \gamma_\beta \gamma_\nu \hat p_2 \gamma^T_\alpha
\frac{(x_1-x_2)\hat p_1}{(x_1-x_2)ys + i\epsilon}
\biggr] B^V(x_1,x_2) \, \varepsilon_{\beta\alpha S^T p_1} \, ,
\nonumber
\end{eqnarray}
where we used $\ell^+\gamma^-=(x_2-x_1)\hat p_1$ and
\begin{eqnarray}
\label{parV2DY}
&&\langle p_2| \bar\psi(\lambda\tilde n^*) \,\gamma_{\mu}\, \psi(0)
| p_2\rangle
\stackrel{{\cal F}_1^{-1}}{=}
p_{2\,\mu } \, \bar q(y) \, .
\end{eqnarray}

We are now in position to check the QED gauge invariance by contraction
with the photon momentum $q_\mu$.
Calculating the trace
\begin{eqnarray}
\label{tr-q}
\frac{1}{4}\, (x_1-x_2)\, \varepsilon_{\beta\alpha S^T p_1} \,
\text{tr}\biggl[ \hat q \gamma_\beta \gamma_\nu \hat p_2 \gamma^T_\alpha
\hat p_1\biggr] =
\varepsilon_{\alpha p_2 S^T p_1}\, g^T_{\alpha\nu}\, y\, (x_1-x_2)\, s \, ,
\end{eqnarray}
one gets
\begin{eqnarray}
\label{FacHadTen4}
q_\mu \, \overline{\cal W}^{(1)}_{\mu\nu}= - \int dx_1 \, dy \,
\biggl[\delta(x_1-x_B) \delta(y-y_B)\biggr] \, \bar q(y)
\varepsilon_{\nu p_2 S^T p_1}\, \int\limits_{-1}^{1} dx_2\, \Im\text{m}
\frac{x_1-x_2}{x_1-x_2+i\epsilon} B^V(x_1,x_2) \not= 0\, ,
\end{eqnarray}
if the gluonic pole is present. Note that here and below we consider
only   the imaginary part of the hadron tensor ( as for any single spin asymmetry).

Let us analyze this problem from a viewpoint of
the so-called $\xi$-process (see \cite{BogoShir}, Section 33.2) applied for 
the partonic sub-process.
Generally speaking, the single diagram on Fig. \ref{Fig-DY}(a) cannot give 
the gauge invariant hadron tensor. One needs the second diagram (cf. ) with
the gluon insertion in the quark line, see Fig. \ref{Fig-DY}(b).

We now focus on the contribution from the diagram depicted on Fig. \ref{Fig-DY}(b).
The corresponding hadron tensor takes the form:
\begin{eqnarray}
\label{HadTen2}
{\cal W}^{(2)}_{\mu\nu}=
\int d^4 k_1\, d^4 k_2 \, \delta^{(4)}(k_1+k_2-q)
\text{tr}\biggl[
\gamma_\mu  {\cal F}(k_1) \gamma_\nu \bar\Phi(k_2)
\biggr]
\, ,
\end{eqnarray}
where the function ${\cal F}(k_1)$ reads
\begin{eqnarray}
\label{PhiF2}
{\cal F}(k_1)= S(k_1) \gamma_\alpha \int d^4\eta_1\, e^{-ik_1\cdot\eta_1}
\langle p_1, S^T | \bar\psi(\eta_1) \, gA^T_{\alpha}(0) \, \psi(0) |S^T, p_1\rangle \, .
\end{eqnarray}
Performing the collinear factorization, we derive the expression for the
factorized hadron tensor which corresponds to the diagram on Fig. \ref{Fig-DY}(b):
\begin{eqnarray}
\label{FacHadTen2}
\overline{\cal W}^{(2)}_{\mu\nu}=  \int dx_1 \, dy \,
\biggl[\delta(x_1-x_B) \delta(y-y_B)\biggr] \, \bar q(y) \,
\text{tr}\biggl[
\gamma_\mu \biggl( \int d^4 k_1\, 
\delta(x_1p_1^+ - k_1^+) {\cal F}(k_1)\biggr) \gamma_\nu \hat p_2 \biggr] \, .
\end{eqnarray}
After some algebra, the integral over $k_1$ in (\ref{FacHadTen2}) can be
rewritten as
\begin{eqnarray}
\label{FacF2}
\int d^4 k_1\, \delta(x_1p_1^+ - k_1^+) {\cal F}^{[\gamma^+]}(k_1)=
\frac{\hat p_2 \gamma_\alpha^T\gamma_\beta}{2p_2^-p_1^+}
\, \varepsilon_{\beta\alpha S^T p_1}\,
\frac{1}{x_1+i\epsilon} \,
\int\limits_{-1}^{1} dx_2\, B^V(x_1,x_2)\, ,
\end{eqnarray}
where the parametrization (\ref{parVecDY}) has been used.
Taking into account (\ref{FacF2}) and calculating the Dirac trace, the
contraction of the tensor $\overline{\cal W}^{(2)}_{\mu\nu}$ with the 
photon momentum $q_\mu$ gives us
\begin{eqnarray}
\label{FacHadTen3}
q_\mu \, \overline{\cal W}^{(2)}_{\mu\nu}= \int dx_1 \, dy \,
\biggl[\delta(x_1-x_B) \delta(y-y_B)\biggr] \,
\bar q(y) \,
 \, \varepsilon_{\nu p_2 S^T p_1}\,
 \int\limits_{-1}^{1} dx_2\, \Im\text{m} \,B^V(x_1,x_2)\, .
\end{eqnarray}
From this, one can observe that if the function $B^V(x_1,x_2)$ is the purely 
real one (see, (\ref{g-pole-B})),
this part of the hadron tensor, which is associated with the diagram on Fig. \ref{Fig-DY}(b),
does not contribute to the imaginary part.

We now study the net effect of the $\overline{\cal W}^{(1)}_{\mu\nu}$ and
$\overline{\cal W}^{(2)}_{\mu\nu}$ contributions and its role for the QED gauge invariance.
adding the contributions of (\ref{FacHadTen4}) and (\ref{FacHadTen3}), one can easily obtain:
\begin{eqnarray}
\label{com}
q_\mu \, \overline{\cal W}^{(1)}_{\mu\nu} + q_\mu \, \overline{\cal W}^{(2)}_{\mu\nu}=
\varepsilon_{\nu p_2 S^T p_1}\, \bar q(y_B)\,
 \Im\text{m}\, \int\limits_{-1}^{1} dx_2\, B^V(x_B,x_2)\,
 \biggl[  \frac{x_B-x_2}{x_B-x_2+i\epsilon} -  1 \biggr]\, .
\end{eqnarray}
If we tacitly assume that $B^V(x_1,x_2)$ is some real and regular (at $x_1 = x_2$) function
that the numerator and denominator in the first term inside the brackets
are contracted and, as a result of this,
both the first and second terms in (\ref{com}) do not have an imaginary part.
That would mean the electromagnetic gauge invariance for the tensor.

The existence of the gluonic pole changes the situation.
Inserting now (\ref{Sol-way-1}) into (\ref{com}), one gets
\begin{eqnarray}
\label{com-2}
q_\mu \, \overline{\cal W}^{(1)}_{\mu\nu} + q_\mu \, \overline{\cal W}^{(2)}_{\mu\nu}=
\varepsilon_{\nu p_2 S^T p_1}\, \bar q(y_B)\,
\Im\text{m}\, \int\limits_{-1}^{1} dx_2\, T(x_B,x_2)\,
 \biggl[  \frac{x_B-x_2}{(x_B-x_2+i\epsilon)^2} -
 \frac{1}{x_B-x_2+i\epsilon}\biggr]\, .
\end{eqnarray}
Performing  the calculation one gets:
\begin{eqnarray}
\label{com-3}
&&q_\mu\, \overline{\cal W}^{(1)}_{\mu\nu} +
q_\mu\, \overline{\cal W}^{(2)}_{\mu\nu} = 0\, .
\end{eqnarray}
This is nothing else than the QED gauge invariance for the imaginary part of the hadron tensor.
From (\ref{com-2}), we can see that the gauge invariance takes place only if the prescriptions
in the gluonic pole and in the quark propagator of the hard part are coinciding.
Indeed Eq. (\ref{com-2}) with the field (\ref{AgAlter}) takes the form
\begin{eqnarray}
\label{com-2-2}
&&q_\mu\, \overline{\cal W}^{(1)}_{\mu\nu} + q_\mu\, \overline{\cal W}^{(2)}_{\mu\nu}=
\nonumber\\
&&\varepsilon_{\nu p_2 S^T p_1}\, \bar q(y_B) \,
\Im\text{m} \,\int\limits_{-1}^{1} dx_2\, T(x_B,x_2)\,
 \biggl[  \frac{x_B-x_2}{(x_B-x_2-i\epsilon)(x_B-x_2+i\epsilon)} -
 \frac{1}{x_B-x_2+i\epsilon}\biggr]\, .
\end{eqnarray}
It is clear that the first term in the brackets is purely real, and the imaginary part from the
second term stays uncompensated. Let us note for completeness, that the treatment of the
pole in the principal value sense is equivalent to the mean arithmetic of two discussed 
prescriptions and also cannot satisfy the gauge invariance. 
Thus we completed the {\it reductio ad absurdum} of the
hint suggested in Section \ref{Causality} and found that the contour gauge (\ref{cg2}) 
is a correct one.
In other words, it means that the prescription in the quark propagator must agree with the
representation of $B^V(x_B,x_2)$. Otherwise, one may face the problem with the gauge invariance.

\begin{figure}[t]
\centerline{\includegraphics[width=0.3\textwidth]{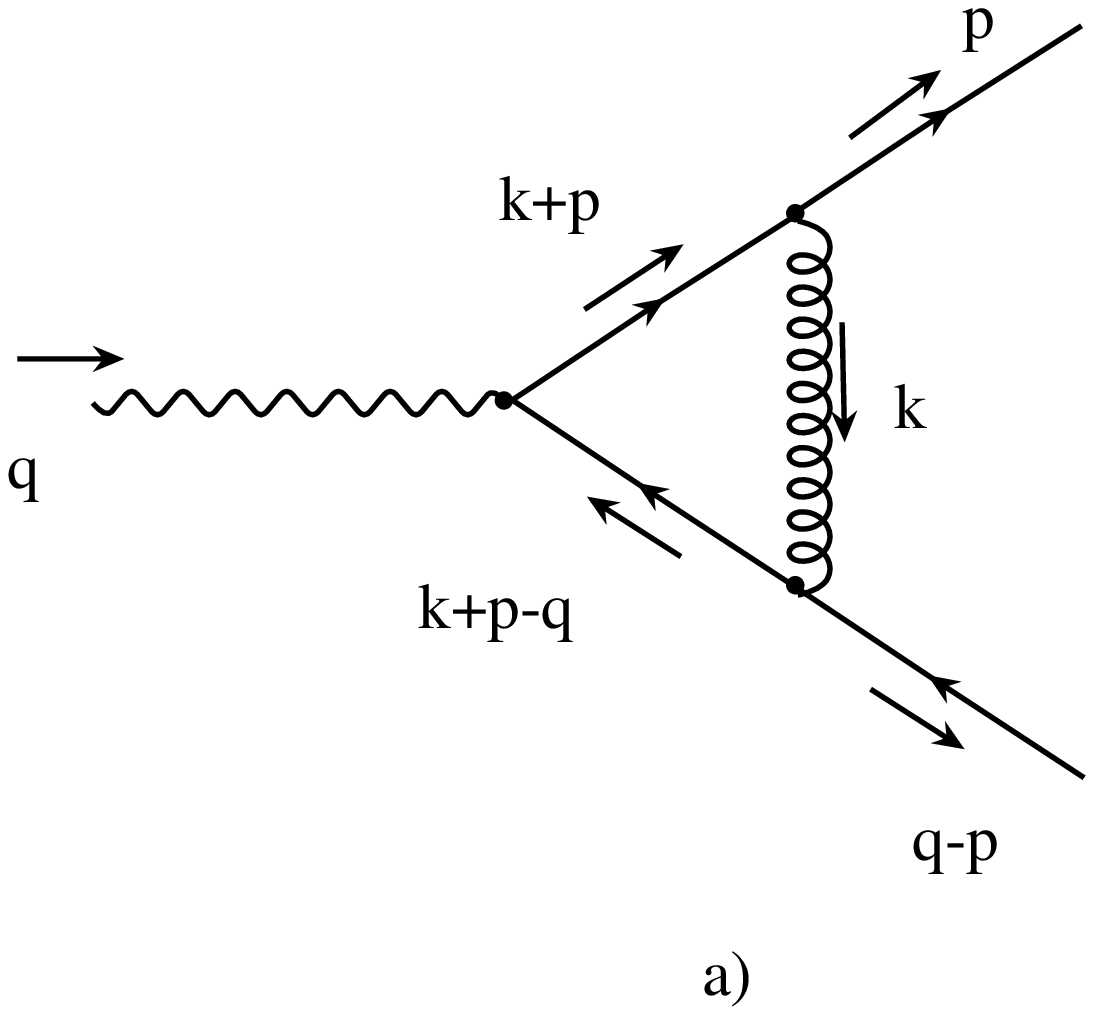}
\hspace{1.cm}\includegraphics[width=0.3\textwidth]{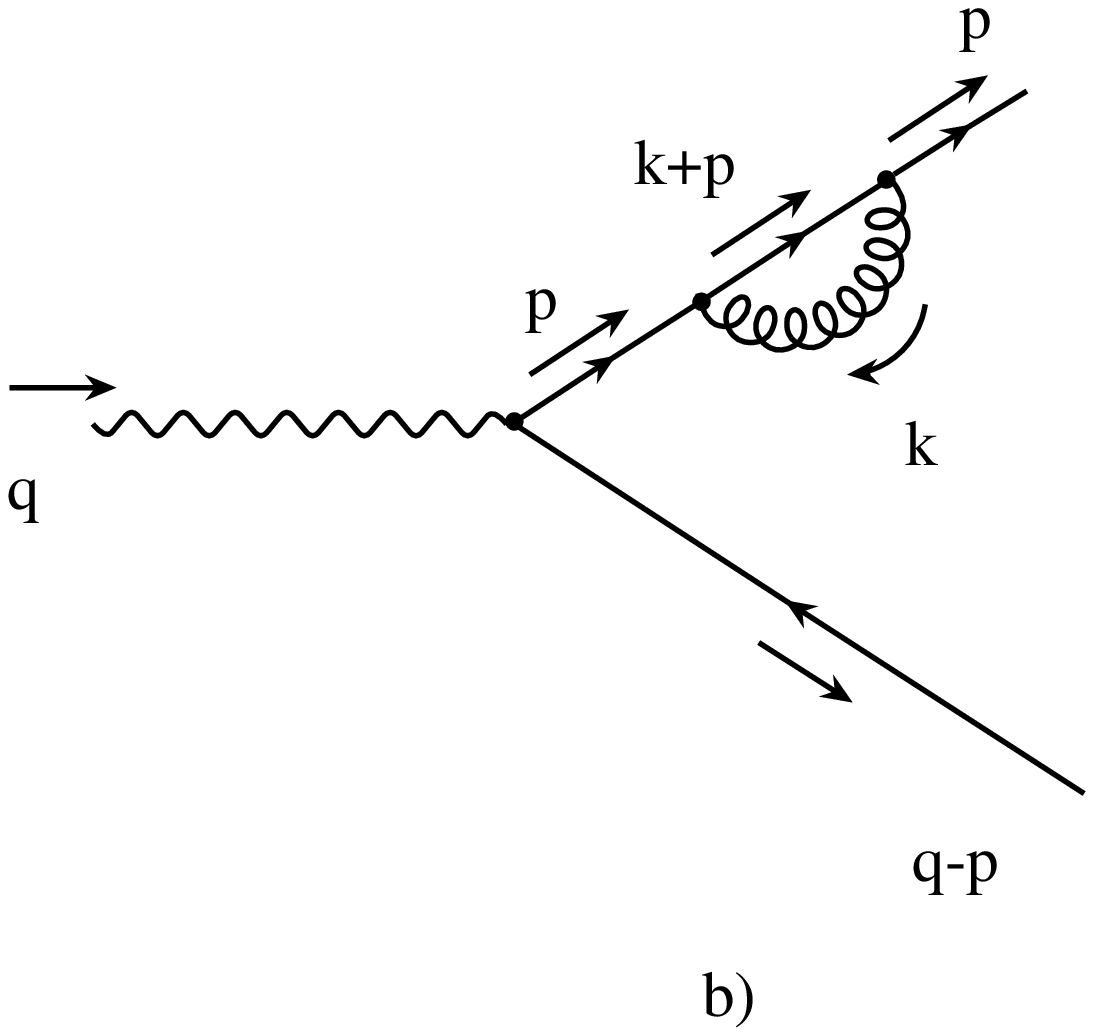}}
\centerline{\includegraphics[width=0.3\textwidth]{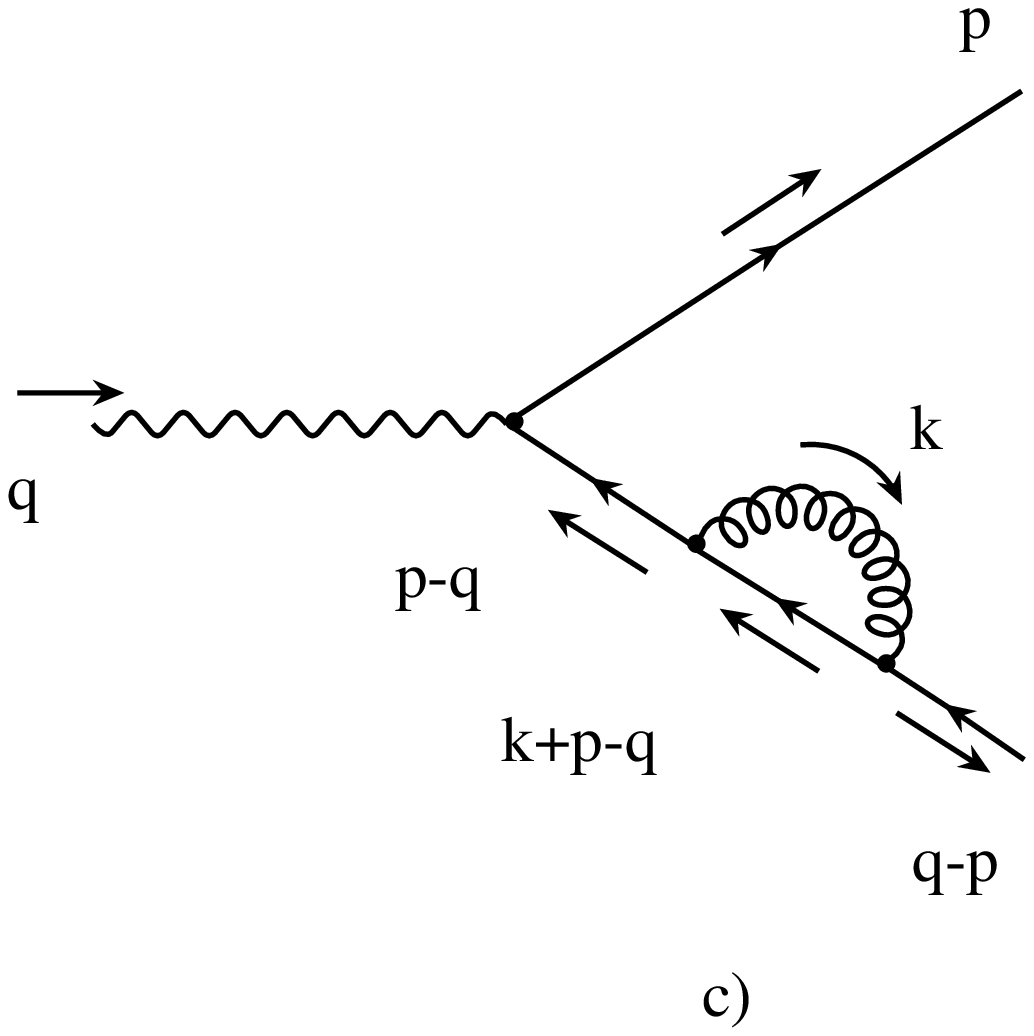}}
\vspace{1cm}
  \caption{The Feynman diagrams which contribute to the $\alpha^3$-order amplitude in QED.}
\label{Fig-QED}
\end{figure}

It is instructive to compare the electromagnetic gauge invariance of the gluonic 
poles contributions with that
of perturbative QCD. In the latter case the imaginary part is provided by hard gluon loops and
the QED gauge invariant set consists of $3$ diagrams depicted on Fig. \ref{Fig-QED}.
At the same time, the imaginary part is due to the
single diagram on Fig. \ref{Fig-QED}(a) and it is gauge invariant by itself 
as the photon line couples to two on-shell
(because of the Cutkosky cutting rule) quarks. This reasoning, however, 
does not happen to work for (non-perturbative)
gluonic pole contribution (see, Fig. \ref{Fig-DY}(a)) and the contribution of
the diagram on Fig. \ref{Fig-DY}(b) should be added to ensure the electromagnetic gauge invariance.
This is clearly seen from Eq. (\ref{com}) where the analog of the contribution of
the diagram on Fig. \ref{Fig-QED}(a) is represented by the first term in the brackets.
Its imaginary part is zero (i.e. QED gauge invariant) only if gluonic pole is absent at all.
This situation corresponds also to the difficulties in the applicability of 
Ward identities to gluonic poles contributions
(see \cite{Koike:2009yb} and Refs. therein).

As we have shown, only the sum of two contributions represented by the diagrams on
Fig. \ref{Fig-DY}(a) and (b) can ensure the electromagnetic gauge invariance.
We now inspect the influence of a ``new" contribution \ref{Fig-DY}(b) on the single spin asymmetry
and obtain the QED gauge invariant expression for the hadron tensor.
It reads
\begin{eqnarray}
\label{HadTen-GI}
\overline{\cal W}^{\text{GI}}_{\mu\nu}=
\overline{\cal W}^{(1)}_{\mu\nu} + \overline{\cal W}^{(2)}_{\mu\nu} =
- \frac{2}{q^2}\,\varepsilon_{\nu S^T p_1 p_2} \, Z_\mu
\, \bar q(y_B)\, T(x_B,x_B) \, ,
\end{eqnarray}
where one used $q^2=s x_B y_B$ and introduced the vector
\begin{eqnarray}
\label{vecZ}
Z_\mu= \widehat p_{1\,\mu} - \widehat p_{2\,\mu} \equiv x_B \, p_{1\,\mu}- y_B\, p_{2\,\mu} \, ,
\end{eqnarray}
which together with the vectors:
\begin{eqnarray}
\label{vecX-Y}
X_\mu= -\frac{2}{s} \biggl[
(Z\cdot p_2)\biggl(p_{1\, \mu} - \frac{q_\mu}{2x_B} \biggr) -
(Z\cdot p_1)\biggl(p_{2\, \mu} - \frac{q_\mu}{2y_B} \biggr)
\biggr], \quad
Y_\mu=\frac{2}{s} \, \varepsilon_{\mu p_1 p_2 q}
\end{eqnarray}
form the mutually orthogonal basis (see, \cite{Barone}).
Here $\widehat p_{i\,\mu}$ are the partonic momenta ($q^\mu=\widehat p_{1\,\mu}+\widehat p_{2\,\mu}$)
With the help of (\ref{vecZ}) and (\ref{vecX-Y}), the lepton momenta can be written as
(this is the lepton c.m. system)
\begin{eqnarray}
\label{lepmom}
l_{1\,\mu} = \frac{1}{2} q_\mu  + \frac{Q}{2} f_\mu(\theta,\varphi; \hat X, \hat Y, \hat Z)\, ,
\quad
l_{2\,\mu} = \frac{1}{2} q_\mu  - \frac{Q}{2} f_\mu(\theta,\varphi; \hat X, \hat Y, \hat Z)\, ,
\end{eqnarray}
where $\hat A = A/\sqrt{-A^2}$ and
\begin{eqnarray}
f_\mu(\theta,\varphi; \hat X, \hat Y, \hat Z)=
\hat X_\mu\, \cos\varphi \,\sin\theta +
\hat Y_\mu\, \sin\varphi \,\sin\theta + \hat Z_\mu\, \cos\theta \, .
\end{eqnarray}
Within this frame, the contraction of the lepton tensor with the gauge invariant
hadron tensor (\ref{HadTen-GI}) reads
\begin{eqnarray}
{\cal L}_{\mu\nu} \, \overline{\cal W}^{\text{GI}}_{\mu\nu} =
-2 \cos\theta\, \varepsilon_{\nu S^T p_1 p_2} \,\bar q(y_B)\, T(x_B,x_B)\, .
\end{eqnarray}
We want to emphasize that this differs by the factor of $2$ in comparison with the case where
only one diagram, presented on Fig. \ref{Fig-DY}(a), has been included in the (gauge non-invariant)
hadron tensor, {\it i.e.}
\begin{eqnarray}
\label{Diff2}
{\cal L}_{\mu\nu} \,\overline{\cal W}^{(1)}_{\mu\nu} =
\frac{1}{2}\, {\cal L}_{\mu\nu} \,\overline{\cal W}^{\text{GI}}_{\mu\nu} \, .
\end{eqnarray}
Therefore, from the practical point of view, the neglecting of the diagram on 
Fig. \ref{Fig-DY}(b) or, in other words,
the use of the QED gauge non-invariant hadron tensor yields the error of the factor of two.

Indeed, taking the contribution of the diagram Fig. \ref{Fig-DY}(a) corresponds to 
keeping of only the term
proportional to $\widehat p_{1\,\mu}$ in (\ref{vecZ}). The contraction with 
(gauge invariant) leptonic tensor is
equivalent to making it gauge invariant by substitution
\begin{eqnarray}
\label{fac2}
\widehat p_{1\,\mu} \Longrightarrow \widehat p_{1\,\mu} - q_\mu \frac{\widehat p\cdot q}{Q^2}
= \frac{p_{1\,\mu} - p_{2\,\mu}}{2} \, .
\end{eqnarray}
It is this factor of $2$ which makes the difference with the correct gauge invariant expression.

\section{Conclusions and Discussions}

The essence of this paper consists in the exploration of the electromagnetic gauge invariance of
the transverse polarized DY hadron tensor. We showed that it is mandatory to include 
a new contribution of
the extra diagram which naively does not have an imaginary part. 
The account for this extra contribution
leads to the amplification of SSA by the factor of $2$.

This new additional contribution emanates
from the complex gluonic pole prescription in the representation of the twist 3 correlator
$B^V(x_1,x_2)$ which, in its turn,
is directly related to the complex pole prescription in the quark propagator forming
the hard part of the corresponding hadron tensor.

In more detail, the causal prescription in the quark propagator, involved in the hard part of
the diagram on Fig. \ref{Fig-DY}(a), selects from the physical axial gauges the contour gauge
defined by Eq. (\ref{cg2}). At the same time, the contour gauge
predestines Eq. (\ref{Ag}) and, therefore, the representation of $B^V(x_1,x_2)$ in
the form of the gluonic pole with the complex prescription, see (\ref{Sol-way-1}).
Since both diagrams on Fig. \ref{Fig-DY}(a) and (b) should be
considered within the same (contour) gauge, the representation (\ref{Sol-way-1}),
which we advocate, has to be applied for the diagram depicted on Fig. \ref{Fig-DY}(b).
As a result of this, the diagram on Fig. \ref{Fig-DY}(b), in contrast to naive assumptions,
has the imaginary part. In some sense, the diagram on Fig. \ref{Fig-DY}(b) feels the
complex prescription in the hard part of the diagram on Fig. \ref{Fig-DY}(a) by means of the contour
gauge which we make used. Note that, from the physical point of view,
the consideration of each of the diagrams on Fig. \ref{Fig-DY} individually makes no sense.

This is completely similar to the case of exclusive dijet production \cite{Braun}
when the pole prescription in (twist two) matrix element of gluonic fields is
controlled by the corresponding hard subprocess.

We have argued that, in addition to the electromagnetic gauge invariance,
the inclusion of new-found contributions corrects by the factor of $2$ the expression for SSA in 
the transverse polarized Drell-Yan process.

Finally,  we proved that the complex prescription in the quark propagator
forming the hard part of the hadron tensor,
the starting point in the contour gauge, the representation of $B^V(x_1,x_2)$ like (\ref{Sol-way-1})
and the electromagnetic gauge invariance of the hadron tensor must be considered together 
as the deeply related items.

\section{Acknowledgements}

We would like to thank A.B.~Arbuzov, D.~Boer, A.V.~Efremov, D.~Ivanov, N.~Kivel, B.~Pire,
M.V.~Polyakov, P.G.~Ratcliffe and N.G.~Stefanis
for useful discussions and comments.
This work is partly supported by the DAAD program and 
the RFBR (grants 09-02-01149 and  09-02-00732).


\end{document}